# Electrically Deformable Liquid Marbles

Edward Bormashenko[1], Roman Pogreb[1], Tamir Stein[1,2], Gene Whyman[1], Marcelo Schiffer[1] and Doron Aurbach[2]

[1]*Ariel University Center of Samaria, Applied Physics Department, P.O.B. 3, Ariel 40700, Israel*

[2]*Bar-Ilan University, Chemistry Department, Ramat-Gan 52900, Israel*

**Short title:** Electrically Deformable Liquid Marbles

Corresponding author's address: Edward Bormashenko, Ariel University Center of Samaria, P.O.B. 3, Ariel 40700, Israel

Phone: +972-3-906-6134

Fax: +972-3-906-6621

E-mail: edward@ariel.ac.il

**Abstract**

Liquid marbles, which are droplets coated with a hydrophobic powder, were exposed to a uniform electric field. It was established that a threshold value of the electric field, 15 cgse, should be surmounted for deformation of liquid marbles. The shape of the marbles was described as a prolate spheroid. The semi-quantitative theory describing deformation of liquid marbles in a uniform electric field is presented. The scaling law relating the radius of the contact area of the marble to the applied electric field shows a satisfactory agreement with the experimental data.

**Introduction**

Liquid marbles, defined as droplets enwrapped with micrometrically scaled hydrophobic particles, have inspired intensive experimental research in the past decade [1–25]. Various powders have been used for manufacturing liquid marbles, including poly(tetrafluoroethylene), polyvinylidene fluoride, polyethylene, lycopodium, aerogels, and others [1–24]. Marbles enwrapped by graphite were reported recently [25]. Liquid marbles are separated from solid or liquid support by "air pockets" in a way similar to Leidenfrost drops [26]. Various applications of marbles were discussed: liquid marbles containing magnetic powder could be used for ferrofluidic applications when activated by a magnetic field [7, 10, 21]. Electrowetting of liquid marbles was reported by McHale *et al* [9]. It was also demonstrated that liquid marbles could be used for revealing water surface pollution with organic contaminants [14]. pH-responsive liquid marbles and marbles coated with graphite were reported recently [18, 19]. We demonstrate in our paper that liquid marbles can be activated with an electric field.

**Experimental**

Preparation of liquid marbles based on deionized water and PVDF particles of sphere-like shape of size of 100 nm was made as described in Ref 10, 11, 27. The marble volume was 10 μl. The effective surface tension γ of the PVDF-coated marbles was established [24] as 75 erg/cm$^2$, close to that of pure water.

Marbles were placed on a 1 mm thick microscope slide and exposed to an electric field between two metallic plates (the distance *d* between them was about 6 mm) as shown in Figure 1. Voltage in the range of 0–6 kV was applied (Pasco, model SF-9586). As a result of the influence of a uniform electric field, the marble was deformed as depicted in Figure 2. Changing the field direction did not influence the results. All the measurements were done using a Ramé-Hart goniometer (model 500).

**Results and discussion**

The semi-axes of marbles as well as the diameter of the contact area (see Figure 3) were measured as a function of the electric field. The measurements were performed in equal time intervals of 3 seconds sequentially. As a result, two dependencies are presented. One of them gives eccentricity as a function of the electric field (Figure 4). The typical experimental result in this figure shows that eccentricity remains constant (very close to zero) up to a field strength of 15 cgse and afterwards it begins to grow. (We use the CGSE unit system, the most suitable for electrostatic problems with basic units gram, centimeter, second and vacuum permeability $\varepsilon_0$=1). This could be understood from the following scaling considerations. The radius $R$ of a 10 µl marble is 1.3 mm, which is smaller than the capillary length $l_c = \sqrt{\gamma/\rho g} \approx 2.7$mm ($\rho$ is the liquid density). Thus, the gravitational energy of the marble is lower than its surface energy. It can be expected that the marble will be markedly deformed by electric field $E$ when the electrostatic energy of the conducting marble $W_{el} \approx VE^2 = 4\pi R^3 E^2/3$ becomes comparable to its surface energy $W_s = 4\gamma\pi R^2$. Equating $W_{el} = W_s$ yields $E^* \sim \sqrt{\gamma/R}$; substitution of $\gamma$ = 75 erg/cm$^2$ and $R$ = 0.13 cm supplies a rough estimation of $E^*$=24 cgse, which is in qualitative agreement with an experimental value of 15 cgse. It should be mentioned that this estimation is very rough and supplies the exaggerated value of $E^*$ (see discussion below).

The second dependence shows the change in diameter of the marble contact area (Figure 5). When the contact radius reaches its minimal value, the marble detaches from the substrate. Measurements in the vicinity of this value become non-reproducible.

To model the behavior of marbles under electric field action, it would be appropriate to consider a *truncated*-prolate-spheroid geometrical form. However, to the best of our knowledge, the electrostatic problem of a truncated spheroid in the uniform electric field has no analytical solution. Therefore, we use the complete-prolate-spheroid model for the geometrical form of marbles for which the electrostatic problem has been solved [28]. In this case the energy of conducting droplet $W$ exerted on electric field $E$ is given by

$$W = mgc + 2\pi a\gamma(a + \frac{c}{e}\arcsin e) - \frac{VE^2 e^3}{(1-e^2)\left(\ln\frac{1+e}{1-e} - 2e\right)}, \quad (1)$$

where $m$ and $V$ are the mass and volume of the droplet, $a$ and $c$ are its short and long semi-axes, respectively, and $\gamma$ is the surface tension. The eccentricity of a prolate spheroid is defined as

$$e = \sqrt{1 - \frac{a^2}{c^2}}. \quad (2)$$

The three terms in (1) present the gravitational, surface, and electrostatic energies. The second term is simply a product of the prolate spheroid surface and marble surface tension. (For derivation of the third term see [28]). Since the mass and volume of the droplet are constant, the total energy in (1) is a function of a single variable, $e$, while

$$c = \left(\frac{3V}{4\pi}\right)^{1/3} (1-e^2)^{-1/3}, \quad (2a)$$

$$a = \left(\frac{3V}{4\pi}\right)^{1/3} (1-e^2)^{1/6} \quad (2b)$$

and can be minimized, e.g., with Mathematica. The numerical values are $m=0.01$ g, $g=980$ cm/s$^2$, $V=0.01$ cm$^3$, and the effective marble surface tension $\gamma=75$ erg/cm$^2$. The results of the calculation are given in Figure 4 and compared there with the experimental data. The qualitative agreement between the calculation and the experimental data can be recognized.

It should be emphasized that for large values of $E$ the electrostatic energy in (1) diverges, and the energy minimum does not exist, that means instability. The stable solution for the deformed droplet exists in the interval $E \sim 10–15$ cgse not very far from the experimental interval $E \sim 15–20$ cgse.

It should be recalled that equation (1) is valid only for a droplet in the form of a complete prolate spheroid having only one common point with a substrate. In reality, a droplet on a substrate has a finite contact area of the radius $r$. This explains the quantitative disagreement of calculations with experimental results depicted in Figure 4.

At the same time, the scaling interrelation between $r$ and the electric field may be found on the basis of the reasoning developed by Aussillous and Quéré [1]. They considered the surface and gravity energies and demonstrated that the requirement of energy minimum leads to the following scaling law:

$$r \sim l_c^{-1} R^2, \qquad (3)$$

where $R$ is the droplet radius and $l_c$ is the capillary length. In the presence of an electric field the last equation may be generalized to

$$r \sim \cdot [\rho(g - F/m)/\gamma]^{1/2} R^2, \qquad (4)$$

where $F$ is an electric force acting on the droplet upwards and effectively decreasing the gravity force. It is obvious that in the first approximation the electrostatic energy of the marble, which is a nonpolar object, in the form of a truncated prolate spheroid is proportional to $\alpha E^2$ where $\alpha$ is the polarizability. Differentiation of this energy relative to the marble elongation leads to the conclusion that the electric force $F$ acting on the marble is also proportional to $E^2$

$$F \sim AE^2, \qquad (5)$$

where $A$ is some constant. The measured and calculated according to (4), (5) dependencies of $r$ on $E$ are compared in Figure 5. The constant $A=0.017$ (in cgse units) was chosen from the requirement $mg = AE^2_{detach}$, i.e., gravity equals the electric force when the marble detaches from a substrate (the experimental value of $E_{detach}$ was about 24 cgse.) Satisfactory agreement of the scaling law given by Formula (4) with the experimental data is recognized ($r$ in (5) was slightly scaled (by 1.1)).

Thus, it seems that the dependence (3) is also confirmed for other uniform fields and not only for gravity. Note that in the present case the droplet volume and mass are conserved ($R$ is constant) but the field strength is changed.

It should be stressed that, in the case of droplets, their shape change under the electric field action is essentially different compared to marbles. In the former case, the dependence of the eccentricity has been established to be linear [29]. Unlike marbles, which are well separated from the substrate by powder particles, liquid droplets are in close (molecular) contact with it. This leads to mutual polarization of both the liquid and the substrate. As a result, the droplet has nonzero dipole moment even in absence of the external electric field. In such a case the energy correction due to the external field, the electric force and eccentricity are linear in $E$ [29].

**Conclusions**

It is demonstrated that liquid marbles could be activated by an electric field. The electric field threshold should be surpassed for deformation of a liquid marble. A semi-quantitative theory describing the behavior of marbles exposed to a uniform electric field is presented. The theory assumes that the shape of the marble could be approximated by a prolate spheroid. The eccentricity of the marble grows non-linearly with the electric field, unlike the case of pure droplets where this dependence is linear [29]. The scaling law relating the radius of the contact area to the applied electric field is presented. The satisfactory agreement of the proposed scaling relation with the experimental data is recognized.

**Acknowledgements**

We are grateful to Professor M. Zinigrad for his generous support of our experimental activity. The authors are grateful to Albina Musin and Yelena Bormashenko for their help in preparing this paper.

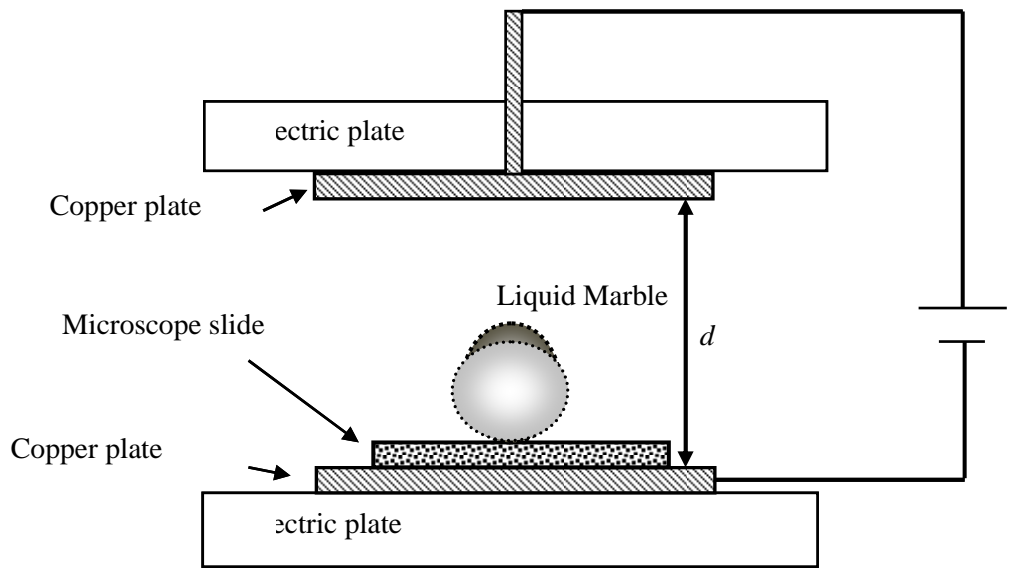

Figure 1. Experimental set for investigation of marble shape under electric field.

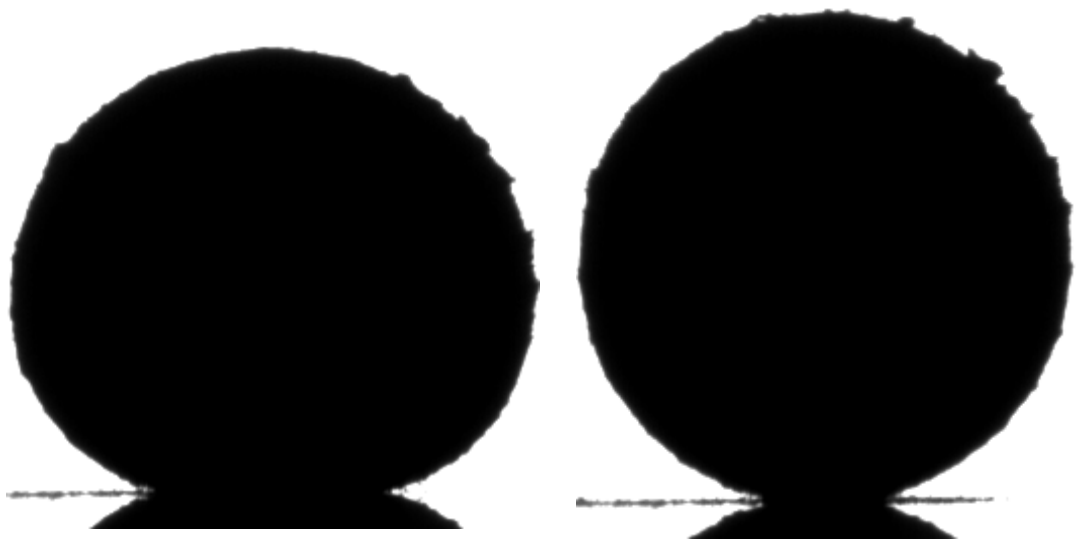

Figure 2. Marble shapes in the absence of electric field (A) and under electric field of 21 cgse (B).

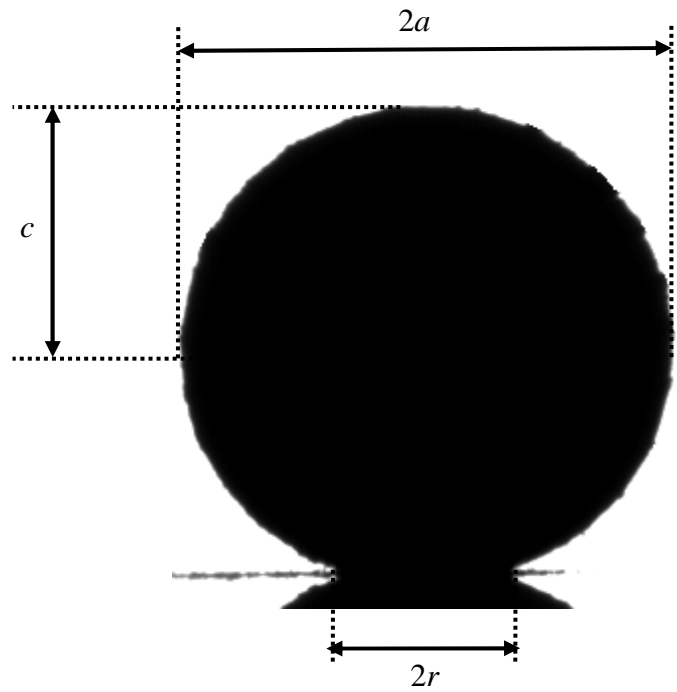

Figure 3. Geometrical parameters of the marble.

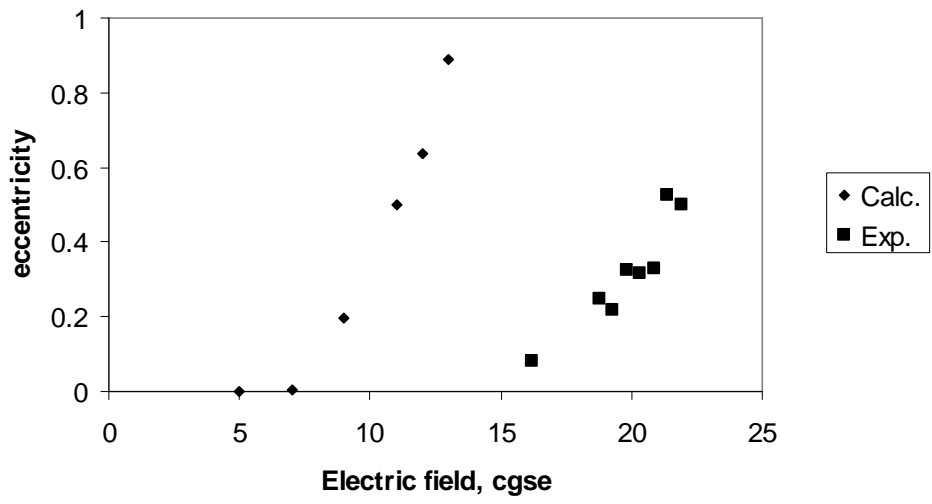

Figure 4. Dependence of the eccentricity of the marble on the applied electric field.

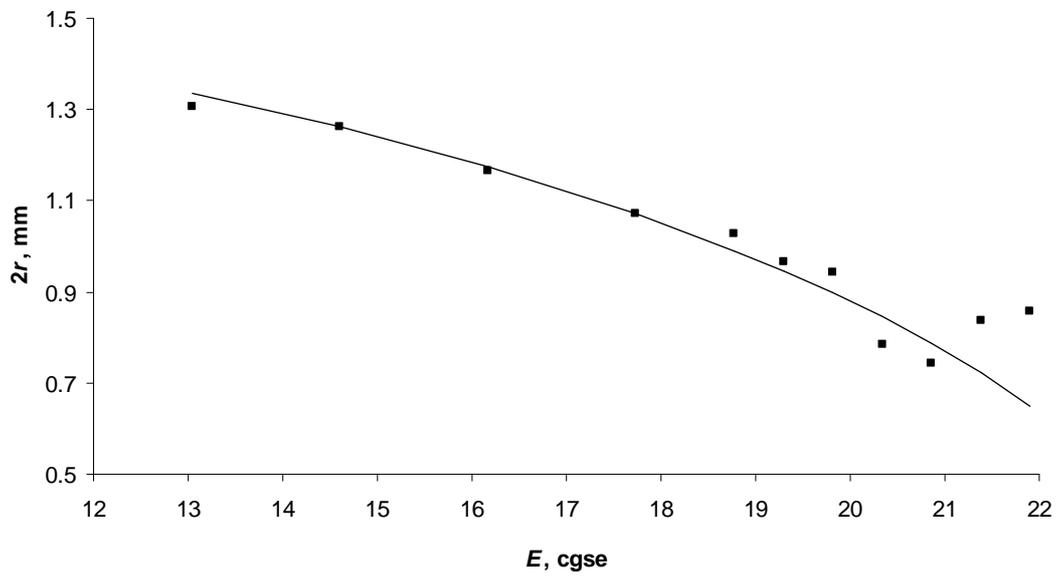

Figure 5. The measured (points) and calculated (curve) according (4) and (1) dependencies of the contact radius on the electric field.